# Beyond esterase-like activity of serum albumin. Histidine-(nitro)phenol radical formation in conversion cascade of p-nitrophenyl acetate and the role of infrared light.

*Magdalena Kowacz\* and Piotr Warszyński*

Jerzy Haber Institute of Catalysis and Surface Chemistry Polish Academy of Sciences; Niezapominajek 8, PL-30239 Krakow

ABSTRACT: Serum albumin, recognized mainly for its capacity to act as a carrier protein for many compounds, can also actively transform some organic molecules. As a starting point in this study we consider esterase-like activity of bovine serum albumin (BSA) toward p-nitrophenyl acetate (p-NPA). Our results reveal that the reaction goes beyond ester hydrolysis step. In fact the transformation product, p-nitrophenol (p-NP), becomes a substrate for further reaction with BSA in which its nitro group in subtracted and released in the form of $HNO_2$. Spectral data indicate that this cascade of events proceeds through formation of phenoxyl radical via proton coupled electron transport (PCET) between OH group of p-NP and imidazole ring of histidine from the protein. Furthermore, the effect of application of electromagnetic radiation in the infrared range, suggests that this remote physical trigger can support interactions based on PCET mechanism by acting on polarization and mutual alignment of water dipoles serving as effective water wires.

INTRODUCTION

Serum albumin, the most abundant protein in blood plasma, unlike many other plasma proteins in not covered by carbohydrate moiety and therefore, serves as a transporter for variety of endogenous and exogenous compounds (1,2). It is a carrier for metal ions, fatty acids or nitric oxide among others. It also to a large extent defines fate of many drugs or toxic substances by transporting them to the place of their action or biotransformation. Apart from its ability to bind and distribute different compounds, serum albumin can actively participate in their transformation. It can catalyze hydrolysis of a variety of organic molecules such as esters, amides and phosphates (1). Very notably bovine serum albumin preserves its catalytic activity toward hydrolysis of esters even at very high temperatures (as high as 160˚C) and also in a denaturated state (3). There is still a debate, if esterase-like activity of serum albumin can be classified as an enzymatic or rather pseudo-enzymatic process. Threr are findings showing irreversible acetylation of albumin residues in reaction with p-nitrophenyl acetate, while enzymatic turnover requires acetylation as an intermediate state and further deacetylation of albumin to yield final reaction products and restore enzyme activity (4). Yet other studies suggested that after initial activity burst related to formation of stable acetylated adducts, there is subsequent phase in which a real turnover on the catalytic site is observed. In any case it has been recognized that tyrosine (first of all Tyr411 but also Tyr150) is the primary residue responsible for catalytic activity of serum albumin towards esters (2). Deprotonation of the catalytic tyrosine is prerequisite for effective hydrolysis. In fact among serum albumins from different species, those having the lowest ionization constant for tyrosine (in particular human serum albumin followed by bovine serum albumin) are the most effective in the hydrolysis of p-nitrophenyl esters (1). Transport of protons from tyrosine to proton acceptor sites in albumin is therefore a prerequisite for operative catalytic process. It has been recognized that nitrogen from the guanidine group of arginine (Arg410) serves as a proton accepting site, but participation of lysine (Lys414) and histidine (His242) in proton transfer reaction on hydrolysis of esters has also been proposed (2). It is recognized that directed proton transfer can be supported by formation of water wires connecting donor and acceptor sites (5). Furthermore, there are studies indicating that structural correlations of water molecules can be enhanced by electromagnetic radiation, especially in the infrared range (6,7). After transient acetylation of albumin by nucleophilic attack of tyrosine on the substrate, the deacetylation step takes place by reaction with water molecule. Hydrolysis of nitrophenyl esters in particular leads to formation of nitrophenol (1).

p-Nitrophenol is a widespread environmental pollutant, toxic to humans and animals. It can cause damage to the central nervous system, liver or kidney. Repeated exposure might result in blood cells injury and mutagenic effects. Therefore, the effective ways of degradation of p-nitrophenol are still sought and extensively studied. There are two major metabolic oxidation pathways proceeding by removal of nitro group by specific enzymes - monooxygenases and formation of either hydroquinone or benzenetriol as the terminal aromatic intermediates that undergo ring cleavage (8). Apart from biological conversion, p-nitrophenol can be oxidized via chemical treatment with oxidizing agents such as hydrogen peroxide or ozone among others.

Anaerobic biological degradation or catalytic reduction (e.g. with the use of nanoparticles) lead in general to formation of 4-aminophenol (9-11). Aqueous photolysis of nitrophenols by solar UV radiation proceeding via radical formation results in release of nitro group in the form of nitrous acid (12).

In this study we explore the activity of bovine serum albumin toward p-nitrophenyl acetate to show new routes of this interaction going beyond usually considered hydrolysis step and including protein-supported transformation of p-nitrophenol. We also demonstrate the potential of the remote physical trigger – infrared radiation to affect those chemical transformations.

EXPERIMENTAL SECTION

Stock solution of bovine serum albumin (BSA from Sigma, 94158) was prepared in in 20 mM imidazole buffer of pH = 7.4 or for supplementary experiments in 20 mM Tris (tris(hydroxymethyl)aminomethane) buffered saline of pH = 7.4. p-Nitrophenyl acetate (p-NPA from Sigma, N 8130) was dissolved in methanol (Uvasol for spectroscopy, 106002 from Merk) and stored refrigerated. Immediately before use the p-NPA solution was diluted with Milli-Q water (1:100 proportion) to get working solution of concentration of 3.5 mM. In order to verify esterase-like activity of BSA, 100 µl of 3 mM solution of p-NPA was added to 400 µl of the protein solution of 3 µM concentration (unless specified otherwise for particular experiments). Subsequently, the evolution of the absorption band centered around 400 nm was monitored with the UV-Vis spectrophotometer (the UV-1800 from Shimadzu) to follow conversion of p-NPA to p-nitrophenol (p-NP). The evolution of UV-Vis spectra was recorded until they have fully developed and no further changes could be observed (usual reaction time was 5 min.). Presented in this work are those final spectra for the completed reaction. Control experiments, accounting for spontaneous hydrolysis of p-NPA were performed in the same manner, but with buffer solution instead of protein in the reaction media.

In order to access a potential effect of infrared radiation (IR) on BSA - supported transformation of p-NPA, the protein solution was exposed to IR light prior to its mixing with the substrate. After the 10 min. exposure, all experimental procedure was exactly the same as described above. Light Emitting Diode with the emission maximum in the IR range at wavelength $\lambda = 2900$ nm (LED29, Roithner-Lasertechnik) and the full width at half maximum of 350 nm, operated by D-31M driver (Roithner-Lasertechnik) in a quasi-continuous wave mode (the mode of maximum average optical power from the LED) at 2 kHz and 200 mA was used as the source of IR light. The emitted IR light represents a non-ionizing electromagnetic radiation (strongly absorbed by liquid water) and its maximum optical power ranged to 29.47 mW. During irradiation samples were thermostated at 22 ˚C with the use of Echotherm IC20 Dry Bath (Torrey Pines Scientific) and their temperature was constantly monitored with TW2 microprobe thermometer (ThermoWorks) by inserted micro thermocouple. To address effect of temperature, additional experiments were performed with samples equilibrated at 40 ˚C before contacting the reactants.

For particular experiments, samples of protein solution were initially degassed with the use of vacuum pump (Buchi, V-100) set at 50 mbar for 1 hour. Samples were then exposed to IR in sealed cuvettes from IR transparent quartz.

RESULTS AND DISCUSSION

As a result of esterase-like activity of bovine serum albumin (BSA) p-nitrophenyl acetate (p-NPA) is converted to p-nitrophenol (p-NP) (Scheme 1.). The transformation is usually evidenced by monitoring evolution of the UV-Vis band centered around 400 nm, which corresponds to the absorption of nitrophenolate ion (yellow at pH values above its pKa = 7.08 at 22 °C) (13).

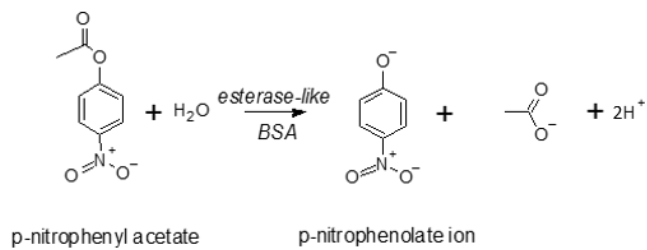

**Scheme 1**. Bovine serum albumin (BSA) – assisted hydrolysis of p-nitrophenyl acetate

In this study we have observed that with increasing concentration of BSA in the reaction mixture, the band position is significantly shifted toward lower wavelengths as shown in Figure 1.

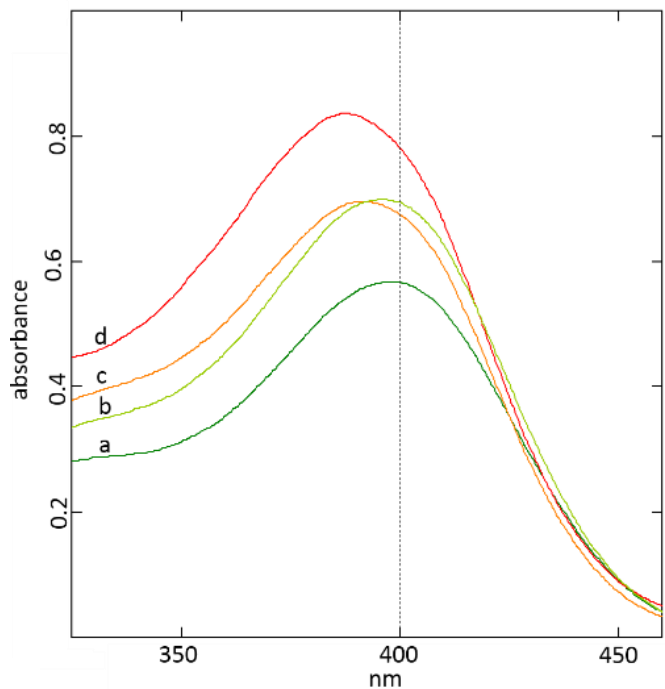

**Figure 1**. Shift in the UV-Vis absorption spectra developing on reaction of p-NPA with BSA of increasing concentration of the protein: a) 0.075 µM; b) 0.375 µM; c) 1.5 µM and d) 3.0 µM.

The second derivative of the UV-Vis spectra illustrated in Figure 2 reveals that the shift is due to the evolution of discrete absorption features that compose the band. New absorption peaks, that appear on increasing protein concentration, are centered at 386 nm, 371 nm, 358 nm and 347 nm. This tetrad of peaks is the characteristic fingerprint of nitrous acid (HNO2) (14,15). Then, fine structure of the band corresponding to evolution of p-nitrophenol indicates the presence of nitrophenolate (400 nm) as well as aci-nitro anion (406 nm) (16). The aci-nitro isomer of nitrophenol is recognized to be an intermediate in the formation of $HNO_2$ in photolysis reaction of nitrophenol (17,18). Therefore, the spectral features observed in our experiments indicate that reaction of BSA and p-NPA goes beyond usually considered hydrolysis step. In fact they strongly point towards possibility of further removal of NO2 group from p-NP and formation of the $HNO_2$.

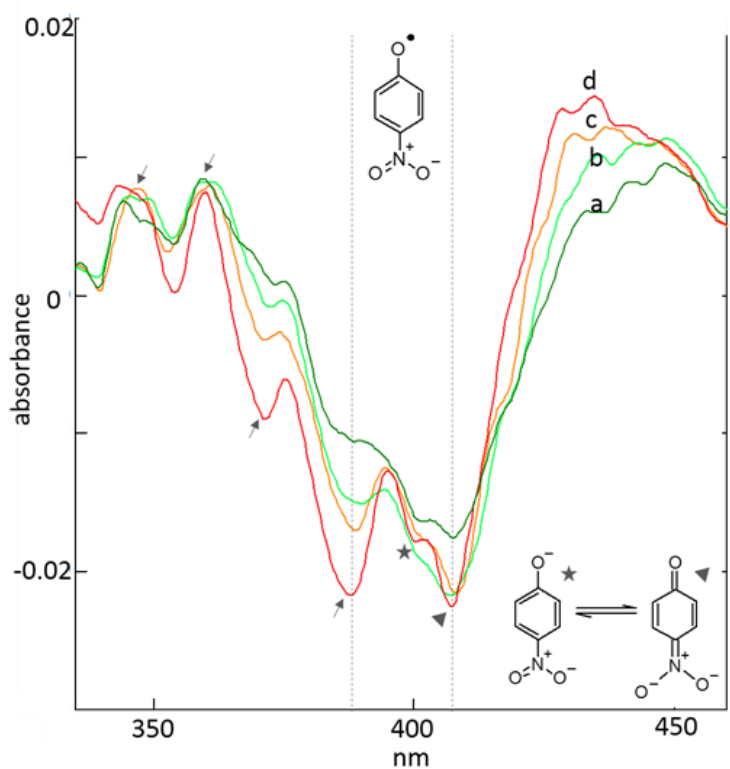

**Figure 2**. Second derivative of the UV-Vis spectra evolving on reaction of p-NPA with BSA of increasing protein concentration of: a) 0.075 µM; b) 0.375 µM; c) 1.5 µM and d) 3.0 µM. Spectral features ascribed to particular species have been marked with arrows for $HNO_2$; star and triangle for p-nitrophenolate anion and its aci-nitro isomer, respectively and dashed lines for p-nitro-phenoxyl radical.

    The mechanism proposed in the literature for the photolysis of NP in aqueous solution involves light-induced breaking of O−H bond and formation of phenoxyl radical, which can yield $HNO_2$ in further reaction of nitro group with water (12). In fact, a detailed inspection of the spectral signatures evolving in our experimental system suggests formation of phenoxyl radical (Figure 2). Its presence is marked by characteristic pair of the UV-Vis peaks of similar intensity located around 386 nm and 406 nm (19-21). It has been stated previously that 386 nm peak belongs to the spectral fingerprint of the $HNO_2$. Yet, this should not be dominant peak in the spectra of $HNO_2$, where its intensity is recognized to be significantly lower than those at 358 nm and 371 nm (14,15). Therefore, the evolution of that strong absorption peak centered around 386 nm accompanied by the absorption band lying in the region assigned to NP-based anions (in which hydrogen has been subtracted from phenol group) can be ascribed to phenoxyl radical formation. A reversible one-electron oxidation of phenol group to yield fairly stable and long-lived phenoxyl radical was observed in the systems where phenol can form OH⋯N intermolecular hydrogen bond with imidazole group (21). Such situation is very possible in our experimental system, where p-NP (a product of esterase-like activity of BSA) can further interact with histidine residues of the protein. It has been shown that phenol oxidation can be reversible exactly due to spontaneous proton transfer from the OH group to imidazole-containing base (22,23). Induction of such transient

phenoxyl radical in our experimental system could enable elimination of the nitro group and release of HNO₂ in further reaction of the radical with water (12). Substitution of NO2 by OH would yield hydroquinone as the oxidation product of p-NP (Scheme 2).

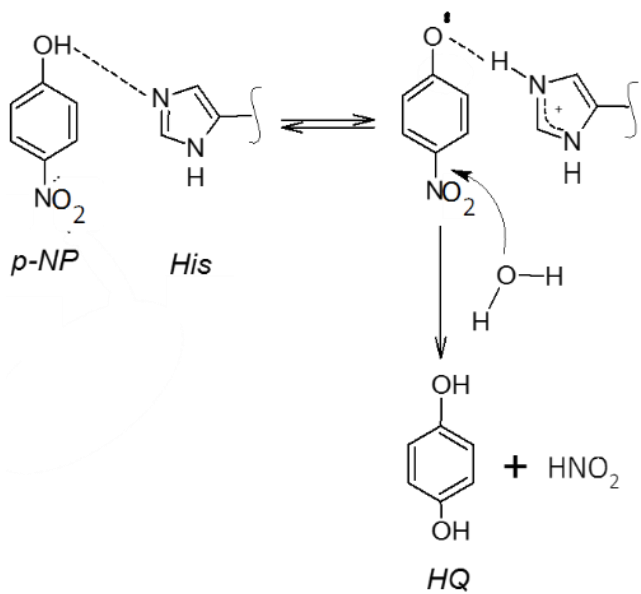

**Scheme 2**. Proposed mechanism of interaction of p-nitrophenole (p-NP) with histidine residue (His) of BSA to yield hydroquinone (HQ) and HNO₂ via nitro-phenoxyl radical intermediate.

Since our experiments were performed predominantly in imidazole buffer that can contribute to the described above mechanism, in order to verify if imidazole group of histidine from BSA is itself sufficient to support phenoxyl radical formation and its further reactivity leading to removal of NO2 group, we changed to TRIS as the buffer medium. The absorption intensity (conversion rate) as well as all spectral signatures of the reaction products were virtually the same as those in imidazole buffer as it is illustrated by curve a) in Figure 3. In the control conditions (without the presence of BSA) only a minor spontaneous hydrolysis of p-NPA can be observed. Yet, second derivative of the control UV-Vis spectra observed in imidazole buffer already bears some resemblance to those generated in BSA-supported reaction (Fig. 3 b). That suggests that the possibility of proton transfer from phenolic OH to imidazole group of the buffer, already may favor some degree of nitrophenol transformation.

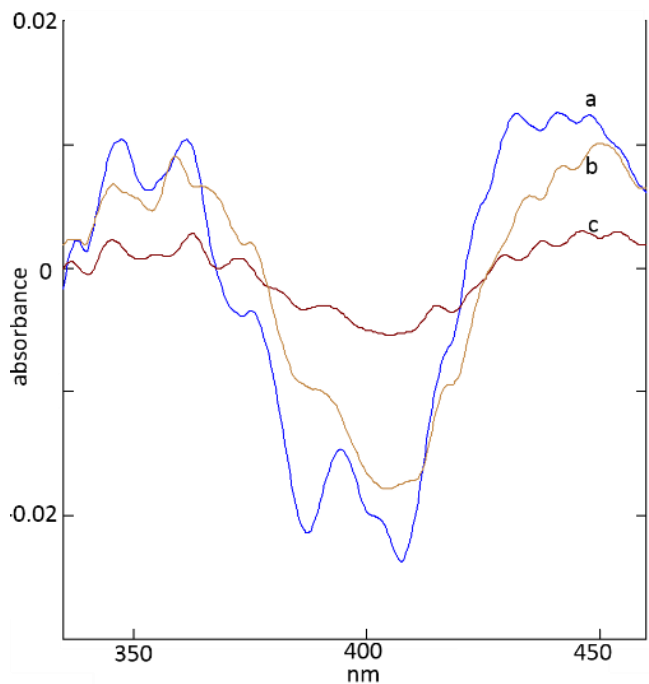

**Figure 3**. Second derivative of the UV-Vis spectra developing on a) interaction of p-NPA with BSA in TRIS buffer and on spontaneous hydrolysis of p-NPA in b) imidazole and c) TRIS buffer media.

Very notably generation of phenoxyl radical by the phenol-imidazole proton coupled electron transfer is a process of great biological importance (23). One-electron transient oxidation of tyrosine (possessing phenol group) to tyroxyl radical accompanied by protonation of imidazole group of histidine enables transfer of electrons over long distances in the photosynthesis or synthesis of DNA. Tyrosine radical-based charge transfer plays also a pivotal role in many other metabolic processes. The electron transfer in general is crucial for energy conversion in cells (24,25). In this context, it has been recognized that water molecules can form structural and functional bridges between OH groups and N atoms in proteins (26). Such bridges (water wires) can conduct protons and effectively support proton coupled electron transfer mechanism. Due to this effect, water wires were recognized to be often essential for enzyme efficiency (27,28). In fact, it has been shown that water molecules are as crucial for proton transport and biological functions of proteins as are amino acids (5).

Our previous studies showed that electromagnetic radiation in the infrared (IR) range had the ability to enhance alignment of water molecules near protein interfaces, by increasing strength and cooperativeness of H-bonds (7). That further affects protein-protein and protein-surface interactions (7,29). The UV-Vis spectra presented in Figure 4 indicate that also in the reaction of BSA with p-NPA, IR light enhances conversion of the latter to the reaction products: p-NPA and $HNO_2$ via phenoxyl radical intermediate. The reaction products were identified on the basis of second derivative of the UV-Vis spectra, with spectral features identical to those generated in BSA-containing systems without irradiation.

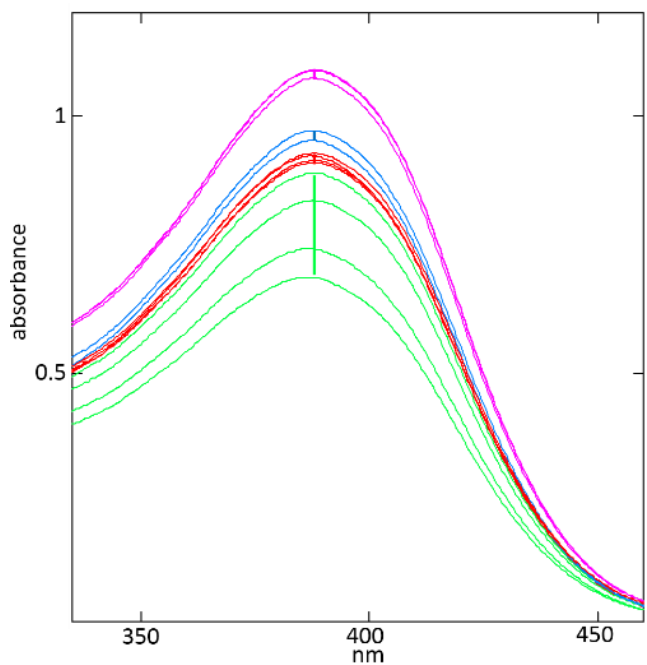

**Figure 4**. UV-Vis spectra developing on interaction of p-NPA with BSA. Color-coded and bar bonded curves represent from bottom to top: light green - standard conditions (no external trigger); red – samples exposed to IR; blue – degassed samples; pink – samples exposed to IR after degassing step.  Lines of the same color denote different experimental batches for the same experimental conditions.

As the IR light of the applied wavelength is only a physical trigger and there is no photochemistry involved, we suggest that the reported effect is related specifically to ability of IR to support water structuring (6,7). Augmented cooperativeness of H-bonds should effectively aid the proton transfer mechanism. Directed proton transfer was recognized to be essential for esterase-like activity of BSA (2) as well as for induction of phenoxyl radical (23). Furthermore, the cooperativeness means more polarized bonds and water dipoles resulting in higher nucleophilic potential of the latter and the nucleophilic attack of water is involved in both, the deacetylation of BSA on hydrolysis of p-NPA (1) as well as in the removal of nitro group from p-NP (12). Bulk thermal effects of IR light can be excluded as monitored sample temperature was changing only by $(0.4 \pm 0.3)°C$, which was way too low to measurably affect the reaction rates. Changes in the conversion efficiency induced by IR light at nearly isothermal conditions were comparable with those promoted by temperature increase from 25°C to 40°C in our experimental setup. To corroborate further the hypothesis of the effect of IR light based on its ability to  support water wires formation, we have performed experiments in degassed solutions (Figure 4).  It has been recognized that proton mobility in solution is greatly increased on the removal of gas cavities that otherwise disrupt the linear chains of water molecules responsible for proton conducting effect (the Grotthus mechanism) (30,31). In fact weak electromagnetic radiation was suggested to act on nanobubbles and to support their escape from the solution bulk that would result in enhanced water structuring (32-34). Therefore, higher reaction yield on degassing and on IR exposure suggests that facilitated proton transport could be due to the common underlying mechanism responsible for the observed effect in both cases. Moreover, at standard experimental conditions (without IR and degassing) significant variance in the reaction yield can be observed between different reaction batches (see Figure 4, bottom, green curves). That is in clear contrast to the narrow distribution of conversion rates for irradiated or degassed samples. Such effect suggests some random factor affecting the reaction at standard conditions and supports the idea of the influence of gas cavities. The amount of gas cavities and their distribution can vary from sample to sample, yet degassing or IR light remove those random distortions. However, exposure to IR light enhances conversion even in previously degassed system (Figure 4, top, pink curves). This can indicate that, in agreement with other studies, water structuring effect of electromagnetic radiation depends also on its ability to polarize molecular dipoles enhancing by this mean their mutual correlations (6,35).

CONCLUSIONS AND PERSPECTIVES

Our results demonstrate that BSA, in addition to its recognized esterase-like activity as evidenced by the hydrolysis of p-NPA to p-NP, can also promote release of the nitro group from the latter. NO2 is eliminated in the form of $HNO_2$ in the reaction proceeding via formation of phenoxyl radical by the intermolecular proton transfer from OH group of p-NP to imidazole ring of histidine. These findings suggest that the most abundant plasma protein can actively initiate degradation of nitrophenolic compounds, which is recognized to be performed on an enzymatic route. Furthermore, in the presence of phenols blood albumin can possibly contribute to generation of radical species with all consequent chemical events. Then $HNO_2$ released in reaction of BSA with NP in solution rapidly decomposes to nitrogen dioxide and nitric oxide. NO is a signaling molecule synthesized by specialized enzymes with many regulatory functions in physiological and pathological processes. The only recognized non-enzymatic pathway of its formation was reported for highly acidic conditions in disease states. Thus, our results point towards possibility of NO generation in blood plasma after exposure to nitrophenols.

These new recognized aspects of chemistry of BSA can supplement our understanding of its complex physiological functions as well as can be employed in bioengineering applications. Especially in the latter context, demonstrated ability to affect the proton-transfer related interactions by means of physical trigger in the form of infrared light can find practical implications. Moreover, the IR light is part of natural solar radiation and its impact can be considered in protein functioning in living organisms.

AUTHOR INFORMATION

**Corresponding Author**

* e-mail: mkowacz@uw.edu

ACKNOWLEDGMENT

This work was financially supported by National Science Centre through grant FUGA number DEC-2015/16/S/ST4/00465 and through statutory research funds to Jerzy Haber Institute of Catalysis and Surface Chemistry, Polish Academy of Sciences

REFERENCES

1. Sakurai, Y.; Ma, S.; Watanabe, H.; Yamaotsu, N.; Hirono, S.; Kurono, Y.; Kragh-Hansen, U.; Otagiri, M. Esterase-Like Activity of Serum Albumin: Characterization of Its Structural Chemistry Using P-Nitrophenyl Esters as Substrates. *Pharmaceutical Res.* **2004**, *21*, 285-292.

2. Goncharov, N.; Belinskaia, D.; Shmurak, V.; Terpilowski, M.; Jenkins, R.; Avdonin, P. Serum Albumin Binding and Esterase Activity: Mechanistic Interactions with Organophosphates. *Molecules* **2017**, *22*, 1201.

3. Córdova, J.; Ryan, J.; Boonyaratanakornkit, B.; Clark, D. Esterase Activity of Bovine Serum Albumin up to 160°C: A New Benchmark for Biocatalysis. *Enzyme and Microbial Technol.* **2008**, *42*, 278-283.

4. Lockridge, O.; Xue, W.; Gaydess, A.; Grigoryan, H.; Ding, S.; Schopfer, L.; Hinrichs, S.; Masson, P. Pseudo-Esterase Activity of Human Albumin. *J. Biol. Chem.* **2008**, *283*, 22582-22590.

5. Gerwert, K.; Freier, E.; Wolf, S. The Role of Protein-Bound Water Molecules in Microbial Rhodopsins. *Biochim. Biophys. Acta - Bioenergetics* **2014**, *1837*, 606-613.

6. Chai, B.; Yoo, H.; Pollack, G. Effect of Radiant Energy on Near-Surface Water. *J. Phys. Chem. B* **2009**, *113*, 13953-13958.

7. Kowacz, M.; Marchel, M.; Juknaité, L.; Esperança, J.; Romão, M.; Carvalho, A.; Rebelo, L. Infrared Light-Induced Protein Crystallization. Structuring of Protein Interfacial Water and Periodic Self-Assembly. *J. Crystal Growth* **2017**, *457*, 362-368.

8. Vikram, S.; Pandey, J.; Bhalla, N.; Pandey, G.; Ghosh, A.; Khan, F.; Jain, R.; Raghava, G. Branching Of The P-Nitrophenol (PNP) Degradation Pathway in Burkholderia Sp. Strain SJ98: Evidences from Genetic Characterization of PNP Gene Cluster. *AMB Express* **2012**, *2*, 30.

9. Uberoi, V.; Bhattacharya, S. Toxicity and degradability of Nitrophenols in Anaerobic Systems. *Water Environment Research* **1997**, *69*, 146-156.

10. Aditya, T.; Pal, A.; Pal, T. Nitroarene Reduction: A Trusted Model Reaction to Test Nanoparticle Catalysts. *ChemComm.* **2015**, *51*, 9410-9431.

11. König, R.; Schwarze, M.; Schomäcker, R.; Stubenrauch, C. Catalytic Activity of Mono- and Bi-Metallic Nanoparticles Synthesized via Microemulsions. *Catalysts* **2014**, *4*, 256-275.

12. Barsotti, F.; Bartels-Rausch, T.; De Laurentiis, E.; Ammann, M.; Brigante, M.; Mailhot, G.; Maurino, V.; Minero, C.; Vione, D. Photochemical Formation of Nitrite and Nitrous Acid (HONO) upon Irradiation of Nitrophenols in Aqueous Solution and in Viscous Secondary Organic Aerosol Proxy. *Environ. Sci. Technology* **2017**, *51*, 7486-7495.

13. Pliego, J.; Mateos, J.; Rodriguez, J.; Valero, F.; Baeza, M.; Femat, R.; Camacho, R.; Sandoval, G.; Herrera-López, E. Monitoring Lipase/Esterase Activity by Stopped Flow in a Sequential Injection Analysis System Using P-Nitrophenyl Butyrate. *Sensors* **2015**, *15*, 2798-2811.

14. van Faassen, E.; Vanin, A. Nitric Oxide Radicals And Their Reactions. *Radicals for Life* **2007**, 3-16.

15. Anastasio, C.; Chu, L. Photochemistry of Nitrous Acid (HONO) and Nitrous Acidium Ion (H2ONO+) in Aqueous Solution and Ice. *Env. Sci. Technology* **2009**, *43*, 1108-1114.

16. Corrie, J.; Katayama, Y.; Reid, G.; Anson, M.; Trentham, D.; Sweet, R.; Moffat, K. The Development and Application of Photosensitive Caged Compounds to Aid Time-Resolved Structure Determination of Macromolecules. *Phil. Trans. Royal Soc. A: Math., Phys. Eng. Sci.* **1992**, *340*, 233-244.

17. Nagaya, M.; Kudoh, S.; Nakata, M. Infrared Spectrum and Structure of the Aci-Nitro Form of 2-Nitrophenol in a Low-Temperature Argon Matrix. *Chem.Phys. Letters* **2006**, *427*, 67-71.

18. Vereecken, L.; Chakravarty, H.; Bohn, B.; Lelieveld, J. Theoretical Study on the Formation of H- and O-Atoms, HONO, OH, NO, and NO2 from the Lowest Lying Singlet and Triplet States Inortho-Nitrophenol Photolysis. *Intl. J. Chem. Kinetics* **2016**, *48*, 785-795.

19. Oliver, T.; Zhang, Y.; Roy, A.; Ashfold, M.; Bradforth, S. Exploring Autoionization and Photoinduced Proton-Coupled Electron Transfer Pathways of Phenol in Aqueous Solution. *J. Phys. Chem. Letters* **2015**, *6*, 4159-4164.

20. Maki, T.; Araki, Y.; Ishida, Y.; Onomura, O.; Matsumura, Y. Construction of Persistent Phenoxyl Radical with Intramolecular Hydrogen Bonding. *J. Am. Chem. Soc.***2001**, *123*, 3371-3372.

21. Benisvy, L.; Blake, A.; Collison, D.; Stephen Davies, E.; David Garner, C.; McInnes, E.; McMaster, J.; Whittaker, G.; Wilson, C. A Phenol–Imidazole Pro-Ligand That can Exist as a Phenoxyl Radical, Alone and When Complexed to Copper(II) and Zinc(II). *Dalton Trans.* **2003**, 1975-1985.

22. Markle, T.; Rhile, I.; DiPasquale, A.; Mayer, J. Probing Concerted Proton-Electron Transfer in Phenol-Imidazoles. *Proc. Natl. Acad. Sci. U.S.A.* **2008**, *105*, 8185-8190.

23. Barry, B. Reaction Dynamics and Proton Coupled Electron Transfer: Studies of Tyrosine-Based Charge Transfer in Natural and Biomimetic Systems. *Biochim. Biophys. Acta - Bioenergetics* **2015**, *1847*, 46-54.

24. Kaila, V. Long-Range Proton-Coupled Electron Transfer in Biological Energy Conversion: Towards Mechanistic Understanding of Respiratory Complex I. *J. Royal Soc. Interface* **2018**, *15*, 20170916.

25. Hammesschiffer, S.; Hatcher, E.; Ishikita, H.; Skone, J.; Soudackov, A. Theoretical Studies of Proton-Coupled Electron Transfer: Models and Concepts Relevant to Bioenergetics. *Coordination Chem. Rev.* **2008**, *252*, 384-394.

26. Utas, J.; Kritikos, M.; Sandström, D.; Åkermark, B. Water as a Hydrogen Bonding Bridge Between a Phenol and Imidazole. A Simple Model for Water Binding in Enzymes. *Biochim. Biophys. Acta - Bioenergetics* **2006**, *1757*, 1592-1596.

27. Bjerregaard-Andersen, K.; Sommer, T.; Jensen, J.; Jochimsen, B.; Etzerodt, M.; Morth, J. A Proton Wire and Water Channel Revealed in the Crystal Structure of Isatin Hydrolase. *J. Biol. Chem.* **2014**, *289*, 21351-21359.

28. Grigorenko, B.; Knyazeva, M.; Nemukhin, A. Analysis of Proton Wires in the Enzyme Active Site suggests a Mechanism of C-Di-GMP Hydrolysis by the EAL Domain Phosphodiesterases. *Proteins: Structure, Function, and Bioinformatics* **2016**, *84*, 1670-1680.

29. Kowacz, M.; Warszyński, P. Effect of Infrared Light on Protein Behavior in Contact with Solid Surfaces. *Colloids Surf. A: Physicochem. Eng. Asp.* **2018**, *557*, 94-105.

30. Pashley, R.; Francis, M.; Rzechowicz, M. The Hydrophobicity of Non-Aqueous Liquids and Their Dispersion in Water under Degassed Conditions. *Curr. Opin. Colloid Interface Sci.* **2008**, *13*, 236-244.

31. Grancha, T.; Ferrando-Soria, J.; Cano, J.; Amorós, P.; Seoane, B.; Gascon, J.; Bazaga-García, M.; Losilla, E.; Cabeza, A.; Armentano, D.; Pardo E. Insights into the Dynamics of Grotthuss Mechanism in a Proton-Conducting Chiral Biomof. *Chem. Mat.* **2016**, *28*, 4608-4615.

32. Shatalov, V. Mechanism of the Biological Impact of Weak Electromagnetic Fields and the in Vitro Effects of Blood Degassing. *Biophysics* **2012**, *57*, 808-813.

33. Vallée, P.; Lafait, J.; Legrand, L.; Mentré, P.; Monod, M.; Thomas, Y. Effects of Pulsed Low-Frequency Electromagnetic Fields on Water Characterized by Light Scattering Techniques: Role of Bubbles. *Langmuir* **2005**, *21*, 2293-2299.

34. Vallée, P. Action of Pulsed Low Frequency Electromagnetic Fields on Physicochemical Properties of Water: Incidence on Its Biological Activity. *Eur. J. Water Qual.* **2006**, *37*, 221-232.

35. Sharma, M.; Resta, R.; Car, R. Dipolar Correlations and the Dielectric Permittivity of Water. *Phys. Rev. Letters* **2007**, *98*.

Table of Contents Image

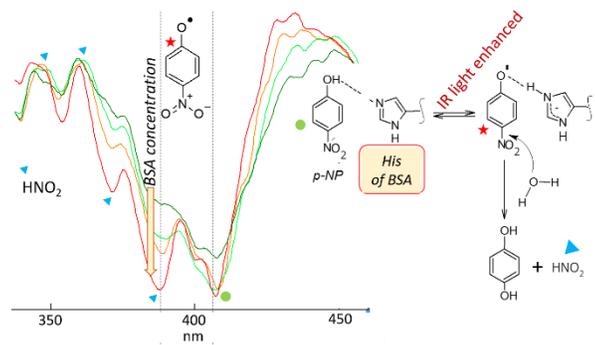